\documentclass[preprint,showpacs, aps]{revtex4}
\usepackage{graphicx}
\begin{document}
\title{Spin-orbit coupling effect on the persistent currents in mesoscopic ring with an Anderson impurity}
\author {Guo-Hui Ding and Bing Dong}
\affiliation{Department of Physics, Shanghai Jiao Tong University,
Shanghai, 200240, China}

\date{\today }

\begin{abstract}
Based on the finite $U$ slave boson method, we have investigated
the effect of Rashba spin-orbit(SO) coupling on the persistent
charge and spin currents in mesoscopic ring with an Anderson
impurity. It is shown that the Kondo effect will decrease the
magnitude of the persistent charge and spin currents in this
side-coupled Anderson impurity case. In the presence of SO
coupling,  the persistent currents change drastically and
oscillate with the strength of SO coupling.  The SO coupling will
suppress the Kondo effect and restore the abrupt jumps of the
persistent currents. It is also found  that a persistent spin
current circulating the ring can exist even without the charge
current in this system.
 \end{abstract}
 \maketitle
\newpage
\section{introduction}
  Recently the spin-orbit(SO) interaction in semiconductor mesoscopic system has
attracted a lot of interest\cite{1}. Due to the coupling of
electron orbital motion with the spin degree of freedom, it is
possible to manipulate and control the electron spin in SO
coupling system by applying an external electrical field or a gate
voltage, and it is believed that the SO effect will play an
important role in the future spintronic application. Actually,
various interesting effects resulting from SO coupling have
already been predicted, such as the Datta-Das spin field-effect
transistor based on Rashba SO interaction\cite{2} and the
intrinsic spin Hall effect\cite{3}.

  In this paper we shall focus our attention  on the persistent charge
current and spin current in mesoscopic semiconductor ring with SO
interaction. The existence of a persistent charge current in a
mesoscopic ring threaded by a magnetic flux has been predicted
decades ago\cite{4}, and has been extensively studied in
theory\cite{5,6,7,8,9} and also observed in various
experiments\cite{10,11,12}. The reason that a persistent charge
current exists may be interpreted as that the magnetic flux
enclosed by the ring introduces an asymmetry between electrons
with clockwise and anticlockwise momentum, thus leads to a
thermodynamic state with a charge current without dissipation. For
a mesoscopic ring with a texture like inhomogeneous magnetic
field, D. Loss et al.\cite{13} predicted that besides the charge
current there are also a persistent spin current. The origin of
the persistent spin current can be related to the Berry phase
acquired when the electron spin precesses during its orbital
motion. The persistent spin current has also been studied in
semiconductor system with Rashba SO coupling term\cite{14,15,16}.
Recently it is shown that a semiconductor ring with SO coupling
can sustain a persistent spin current even in the absence of
external magnetic flux\cite{17}.

For the system of a mesoscopic ring with a magnetic impurity, the
persistent charge current has been investigated in the context of
a mesoscopic ring coupled with a quantum
dot\cite{18,19,20,21,22,23,24}, where the quantum dot acts as an
impurity level and will introduce charge  or spin fluctuations to
the electrons in the ring. The Kondo effect arising from a
localized electron spin interacting with a band of electrons will
be essential in the charge transport in the ring. But to our
knowledge in these systems the SO effect hasn't been considered.
It might be expected that the interplay between the Kondo effect
and the SO coupling in the ring can give new features in the
persistent currents. In this paper we shall address this problem
and investigate the SO effect on persistent charge and spin
currents in the ring system with an Anderson impurity. The
Anderson impurity can act as a magnetic impurity when the impurity
level is in single electron occupied state and as well as a
barrier potential in empty occupied regime.

   The outline of this paper is as follows. In section II we introduce
the model Hamiltonian of the system and also the method of
calculation by finite-U slave boson approach\cite{25,26,27,28}. In
section III  the results of persistent charge current and spin
current are presented and discussed. In Section IV we give the
summary.

\section{Mesoscopic ring with an Anderson impurity}
 The electrons in a closed ring with
SO coupling of Rashba term can be described  by following
Hamiltonian in the polar coordinates\cite{14,29}
\begin{equation}
H_{ring}=\Delta(-i{\partial\over{\partial\varphi}}+{\Phi\over\Phi_0})^2
 +{\alpha_R\over
2}[(\sigma_x\cos\varphi+\sigma_y\sin\varphi)(-i{\partial\over{\partial\varphi}}+{\Phi\over\Phi_0})+h.c.]\;,
\end{equation}
where $\Delta=\hbar^2/(2m_ea^2)$, $a$ is the radius of the ring.
$\alpha_R$ will characterize the strength of Rashba SO
interaction.  $\Phi$ is the external magnetic flux enclosed by the
ring, and $\Phi_0=2\pi\hbar c/e$ is the flux quantum.

We can write the above Hamiltonian in terms of creation and
annihilation operators of electrons in the momentum space,
\begin{equation}
H_{ring}=\sum_{m,\sigma}\epsilon_m
c^\dagger_{m\sigma}c_{m\sigma}+1/2\sum_m[t_m(c^\dagger_{m+1\downarrow}c_{m\uparrow}
+c^\dagger_{m-1\uparrow}c_{m\downarrow})+h.c.]\;,
\end{equation}
where $\epsilon_m=\Delta(m+\phi)^2$, $t_m=\alpha_R
(m+\phi)$,($m=0,\pm 1,\cdots,\pm M$) with $\phi=\Phi/\Phi_0$. One
can see that the SO interaction causes the $m$ mode electrons
coupled with $m+1$ and $m-1$ mode electrons and spin-flip process.
We consider the system with a side-coupled impurity which can be
described by the Anderson impurity model,
\begin{equation}
H_d=\sum_\sigma\epsilon_d d^\dagger_\sigma
d_\sigma+Un_{d\uparrow}n_{d\downarrow}\;.
\end{equation}
The tunneling between the impurity level and the ring are given by
\begin{equation}
H_{d-ring}=t_D\sum_{m\sigma}(d^\dagger_\sigma c_{m\sigma}+h.c)\;.
\end{equation}
 Then the total Hamiltonian for the system should be
\begin{equation}
H=H_{ring}+H_d+H_{d-ring}\;.
\end{equation}

  In order to treat the strong on-site Coulomb interaction in the impurity level.
we adopt the finite-U slave boson approach\cite{25,26}.  A set of
auxiliary bosons $e, p_{\sigma}, d$ is introduced for the impurity
level, which act as projection operators onto the empty, singly
occupied(with spin up and spin down), and doubly occupied electron
states on the impurity, respectively. Then the fermion operators
$d_{\sigma}$ are replaced by $d_{\sigma}\rightarrow
f_{\sigma}z_{\sigma} $, with $z_{\sigma}=e^\dagger
p_{\sigma}+p^\dagger_{\bar\sigma}d$. In order to eliminate
un-physical states, the following constraint conditions are
imposed :$\sum_{\sigma} p^\dagger_{\sigma}p_{\sigma}+e^\dagger
e+d^\dagger d=1$, and
$f^\dagger_{\sigma}f_{\sigma}=p^\dagger_{\sigma}p_{\sigma}+d^\dagger
d(\sigma=\uparrow, \downarrow)$. Therefore,  the Hamiltonian  can
be rewritten as the following effective Hamiltonian in terms of
the auxiliary boson $e, p_{\sigma}, d$  and the pesudo-fermion
operators $f_{\sigma}$:
\begin{eqnarray}
H_{eff}&=&\sum_{m,\sigma}\epsilon_m
c^\dagger_{m\sigma}c_{m\sigma}+1/2\sum_m[t_m(c^\dagger_{m+1\downarrow}c_{m\uparrow}
+c^\dagger_{m-1\uparrow}c_{m\downarrow})+h.c.]
\nonumber\\
&+&\sum_{\sigma}\epsilon_d f^\dagger_{\sigma}f_{\sigma}+
Ud^\dagger d
\nonumber\\
& +&\sum_{m,\sigma } (t_Dz^\dagger_{\sigma}
f^\dagger_{\sigma}c_{m\sigma}+h.c.) + \lambda^{(1)}(\sum_{\sigma}
p^\dagger_{\sigma}p_{\sigma}+e^\dagger e+d^\dagger d-1)
\nonumber\\
&+&\sum_{\sigma}\lambda^{(2)}_{\sigma}(f^\dagger_{\sigma}f_{\sigma}-p^\dagger_{\sigma}p_{\sigma}-d^\dagger
d )\;,
\end{eqnarray}
 where the constraints are incorporated by the Lagrange multipliers $\lambda^{(1)}$ and
$\lambda^{(2)}_{\sigma}$. The first constraint can be interpreted
as a completeness relation of the Hilbert space on the impurity
level, and the second one equates the two ways of counting the
fermion occupancy for a given spin. In the framework of the
finite-U slave boson mean field theory\cite{25,26}, the slave
boson operators $e, p_{\sigma}, d $ and the parameter $z_\sigma$
are replaced by real c numbers. Thus the effective Hamiltonian is
given as
\begin{eqnarray}
H^{MF}_{eff}&=&\sum_{m,\sigma}\epsilon_m
c^\dagger_{m\sigma}c_{m\sigma}+1/2\sum_m[t_m(c^\dagger_{m+1\downarrow}c_{m\uparrow}
+c^\dagger_{m-1\uparrow}c_{m\downarrow})+h.c.]
\nonumber\\
&+&\sum_{\sigma}{\tilde\epsilon_{d\sigma}}f^\dagger_{\sigma}f_{\sigma}
+\sum_{m\sigma } ({\tilde t_{D\sigma}}
f^\dagger_{\sigma}c_{m\sigma}+h.c.)+E_g\;,
\end{eqnarray}
where ${\tilde t_{D\sigma}}=t_Dz_\sigma$ represents the
renormalized tunnel coupling between the impurity and the
mesoscopic ring. $z_\sigma$ can be regarded as the wave function
renormalization factor.
${\tilde\epsilon_{d\sigma}}=\epsilon_d+\lambda^{(2)}_{\sigma}$ is
the renormalized impurity level and $E_g= \lambda^{(1)}(
\sum_{\sigma}
p_{\sigma}^2+e^2+d^2-1)-\sum_{\sigma}\lambda^{(2)}_\sigma(p_{\sigma}^2+d^2)+Ud^2$
is an energy constant.

In this mean field approximation the Hamiltonian is essentially
that of a non-interacting system, hence the single particle energy
levels can be calculated by numerical diagonalization of the
Hamiltonian matrix. Then the ground state of this system
$|\psi_0>$ can be constructed by adding electrons to the lowest
unoccupied energy levels consecutively . By minimizing the ground
state energy with respect to the variational parameters a set of
self-consistent equations can be obtained as in Ref.[27,28], and
they can be applied to determine the variational parameters in the
effective Hamiltonian.

\section{the persistent charge current and spin current }
  In this section we will present the results of our calculation of
the persistent charge current and spin current circulating the
mesoscopic ring. Since there is still some controversial in the
literature for the definition of the spin current operator in the
ring system with SO coupling term\cite{30}. We give both the
formula of charge and spin currents used in this paper explicitly.
It is easy to obtain that the $\varphi$ component of electron
velocity operator in this SO coupled ring is
\begin{equation}
v^\varphi={a\over\hbar}[2\Delta(-i{\partial\over{\partial\varphi}}+\phi)
+\alpha_R(\sigma_x\cos\varphi+\sigma_y\sin\varphi)]\;.
\end{equation}
Thereby the charge current operator is define as $\hat I=-e
v^\varphi$, and in terms of creation and annihilation operator it
can be written as
\begin{equation}
\hat I=-{e a\over \hbar}[2\Delta\sum_{m,\sigma}\
c^\dagger_{m\sigma}c_{m\sigma}(m+\phi)+\alpha_R\sum_m(c^\dagger_{m+1\downarrow}c_{m\uparrow}
+c^\dagger_{m-1\uparrow}c_{m\downarrow})]\;.
\end{equation}
At zero temperature, the persistent charge current  is given by
the  expectation value of the above charge current operator in the
ground state, $I={1\over{2\pi a}}<\psi_0|\hat I|\psi_0>$, and it
can also be calculated from the expression
\begin{equation}
I=-c{\partial E_{gs}\over{\partial\Phi}}=-{e\over
h}<\psi_0|{\partial H\over{\partial\phi}}|\psi_0>\;,
\end{equation}
where $E_{gs}$ is the ground state energy.

In Fig.1 the persistent charge current vs. the enclosed magnetic
flux is plotted for a set of values for the SO coupling strength.
Here we have taken the model parameters $\Delta=0.01, t_D=0.3,
U=2.0$ and the total number of electrons $N$ is around $100$. In
this case one can obtain the Fermi energy of the system $E_F=6.25$
and the level spacing $\delta=0.5$ around the Fermi surface. We
consider the energy level of the Anderson impurity is well below
the Fermi energy( with $\epsilon_d-E_F=-1.0$), therefore the
Anderson impurity is in the Kondo regime. One can see in Fig.1
that the characteristic features of persistent charge current
depends on the parity of the total number of electrons($N$), and
can be distinguished by two cases with $N$ odd and $N$ even. This
is attributed to the different occupation patterns of the highest
occupied single particle energy level in the mean field effective
Hamiltonian. The persistent charge current for the system with
$N+2$ electrons is different from that with $N$ electrons by a
$\pi$ phase shift $I^{N+2}(\phi)=I^{N}(\phi+\pi)$. In case (I)
where the electron number is odd($N=4n-1$ and $N=4n+1$), one
electron is almost localized on the impurity level and forming a
singlet with electron cloud in the conducting ring. This phenomena
leads to the well known Kondo effect. Fig.1 shows that the Kondo
effect decreases the magnitude of the persistent charge current,
and also makes its curve shape resemble sinusoidal.  In the
presence of finite SO coupling($\alpha_R<\Delta$), the spin-up and
spin-down electrons are coupled and it causes the splitting of the
twofold degenerated energy levels in the effective Hamiltonian. It
turns out that the Kondo effect is suppressed and the abrupt jumps
of the persistent charge current with similarity to that of ideal
ring case appears. It is explained in Ref.[14] that the jumps of
the persistent charge current in the case of odd number of
electrons are due to a crossing of levels with opposite spin. In
case (II) where $N$ is even ($N=4n$ and $N=4n+2$), The Kondo
effect is manifested that the magnitude of persistent charge
current is significantly suppressed compared with ideal ring case
and the rounding of the jumps of persistent charge current due to
the level crossing. In the presence of finite SO coupling, the
persistent charge current decreases with increasing the SO
coupling strength when $\alpha_R<\Delta$.

Fig.2 displays the persistent charge current as a function of the
SO coupling strength $\alpha_R$ at different enclosed magnetic
flux. The persistent charge current exhibits oscillations with
increasing the value of $\alpha_R$ for both the systems with even
or odd number of electrons. Therefore by tuning the SO coupling
strength, the magnetic response of this system can change from
paramagnetic to diamagnetic and vice versa. It indicates that SO
coupling can play a important role in electron transport in this
mesoscopic ring. The curve of the persistent charge current for
odd number of electrons shows discontinuity in its derivation,
this can be attributed the level crossing in the energy spectrum
by changing  $\alpha_R$. It is also noted that the position of
this discontinuity for odd $N$ also corresponds to the peak or
valley in even N case.

Since the electron has the spin degree of freedom as well as the
charge, the electron motion in the ring may give rise to a spin
current besides the charge current. Now we turn to study the
persistent spin current in the ground state. The spin current
operator is defined by $\hat J_v=(v^\varphi\sigma_v+\sigma_v
v^\varphi)/2$, which can be written explicitly as
\begin{equation}
\hat
J_v={a\over\hbar}\{2\Delta(-i{\partial\over{\partial\varphi}}+\phi)\sigma_v+{\alpha_R\over
2} [ (\sigma_x\cos\varphi+\sigma_y\sin\varphi)\sigma_v+h.c.]\}\;,
\end{equation}

Therefore the three component of spin current operator in terms of
creation and annihilation operators are given by
\begin{equation}
\hat J_z={a\over\hbar}[2\Delta\sum_{m}\
(c^\dagger_{m\uparrow}c_{m\uparrow}-c^\dagger_{m\downarrow}c_{m\downarrow})(m+\phi)]\;,
\end{equation}

\begin{equation}
\hat
J_x={a\over\hbar}[2\Delta\sum_{m}(c^\dagger_{m\uparrow}c_{m\downarrow}
+c^\dagger_{m\downarrow}c_{m\uparrow})(m+\phi) +{\alpha_R\over
2}\sum_{m,\sigma}(c^\dagger_{m+1\sigma}+c^\dagger_{m-1\sigma})c_{m\sigma}]\;,
\end{equation}

\begin{equation}
\hat J_y={a\over\hbar}[-2i\Delta\sum_{m}(
c^\dagger_{m\uparrow}c_{m\downarrow}-
c^\dagger_{m\downarrow}c_{m\uparrow})(m+\phi) -i{\alpha_R\over
2}\sum_{m,\sigma}(c^\dagger_{m+1\sigma}-c^\dagger_{m-1\sigma})c_{m\sigma}]\;,
\end{equation}
The expectation value of the spin current $J_v={1\over {2\pi
a}}<\psi_0|\hat J_v|\psi_0>$.

In our calculation we find that only the $z$ component of the spin
current is nonzero in the ground state. Fig.3 shows the persistent
spin current $J_z$ vs. magnetic flux at different SO coupling
strength. The persistent spin current is a periodic function of
the magnetic flux $\phi$, which has the even parity symmetry
$J_z(-\phi)=J_z(\phi)$ and also an additional symmetry
$J_z(\phi)=J_z(\pi-\phi)$.  It is noted that the persistent spin
current has quite different dependence behaviors on magnetic flux
compared with the persistent charge current in Fig.1.    In the
presence of finite SO coupling, the persistent spin current is
nonzero both for the systems with odd $N$ and even $N$ at zero
magnetic flux, it indicates that a persistent spin current can be
induced solely by SO interaction without accompany a charge
current. This phenomena is also shown in Ref.[17] where a SO
coupling/normal hybrid ring was considered.

In Fig.4 the persistent spin current $J_z$ as a function of SO
coupling strength is plotted.  In the absence of SO coupling
$\alpha_R=0$, the persistent spin current is exactly zero for both
even and odd number electron system. In the presence of SO
coupling,  The persistent spin current becomes nonzero and shows
oscillations with increasing $\alpha_R$. It can change from
positive to negative values or vice versa by tuning the SO
coupling strength. The sign of the persistent spin current also
shows dependence on the enclosed magnetic flux. For the system
with odd $N$, there is abrupt jumps in the curve of persistent
spin current at certain value of $\alpha_R$, the reason for the
jump is the same as that in the charge current, and is due to the
level crossing in the energy spectrum. It is noted that the
position of the jump coincides with that of the persistent charge
current. This kind of characteristic feature of the persistent
currents might provide a useful way to detect the SO coupling
effects in semiconductor ring system.

\section{conclusions}
 In summary, we have investigated the Rashba SO coupling effect on the persistent charge current
 and spin current in a mesoscopic ring with an Anderson impurity.
 The Anderson impurity leads to the Kondo effect and decreases the
 amplitude of the persistent charge and spin current in the ring. In the semiconducting ring with SO interaction,
 the persistent charge current changes significantly by tuning the SO coupling strength, e.g. from the paramagnetic to
 diamagnetic current. Besides the persistent charge current, there also
 exists a persistent spin current,  which also oscillates with the SO coupling strength. It is shown
 that at zero magnetic flux a persistent spin current can exist even without the charge current.
 Since the persistent spin current can generate an electric field\cite{31}, one might expect
 that experiments on semiconductor ring with Rashba SO coupling can detect the persistent spin current.

\begin{acknowledgments}{ This project is supported by the National Natural Science Foundation of China, the
Shanghai Pujiang Program, and Program for New Century Excellent
Talents in University (NCET). }
\end{acknowledgments}


\begin{figure}[ht]
\includegraphics[width=0.8\columnwidth ]{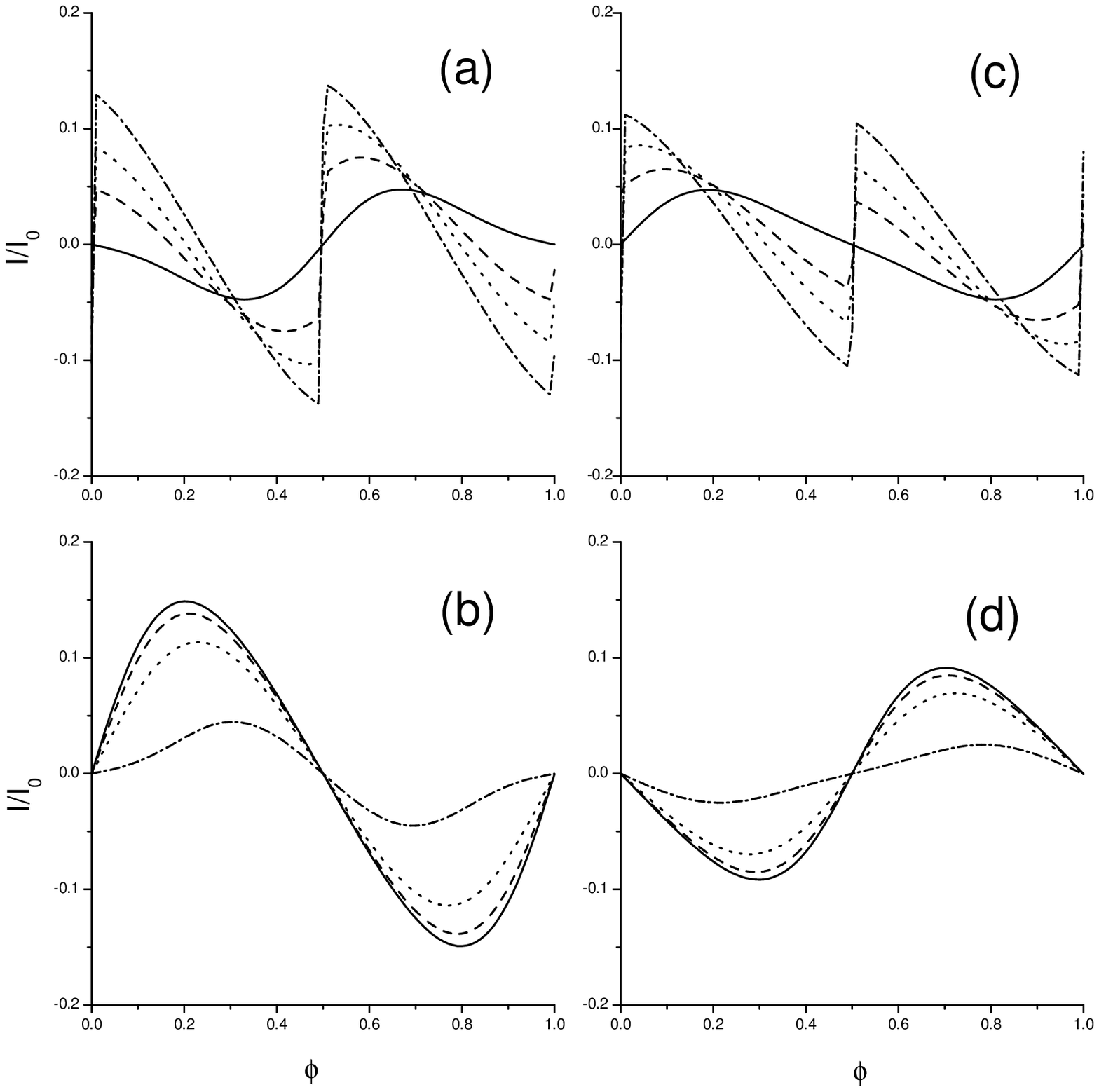}
\caption{ The persistent charge current vs. magnetic flux for a
set of values for the spin-orbit coupling
strength($\alpha_R/\Delta=0.0$(solid line),$ 0.5$(dashed line),
$0.7$(dotted line),$1.0$(dash-dotted line)). The total number of
electrons $N=99$ (a), $100$(b), $101$(c), $102$(d). We take the
other parameters $\Delta=0.01, t_d=0.3, \epsilon_d-E_F=-1.0,
U=2.0$ in the calculation. The persistent charge current is
measured in units of $I_0=e N\Delta $. }
\end{figure}

\begin{figure}[ht]
\includegraphics[width=0.8\columnwidth ]{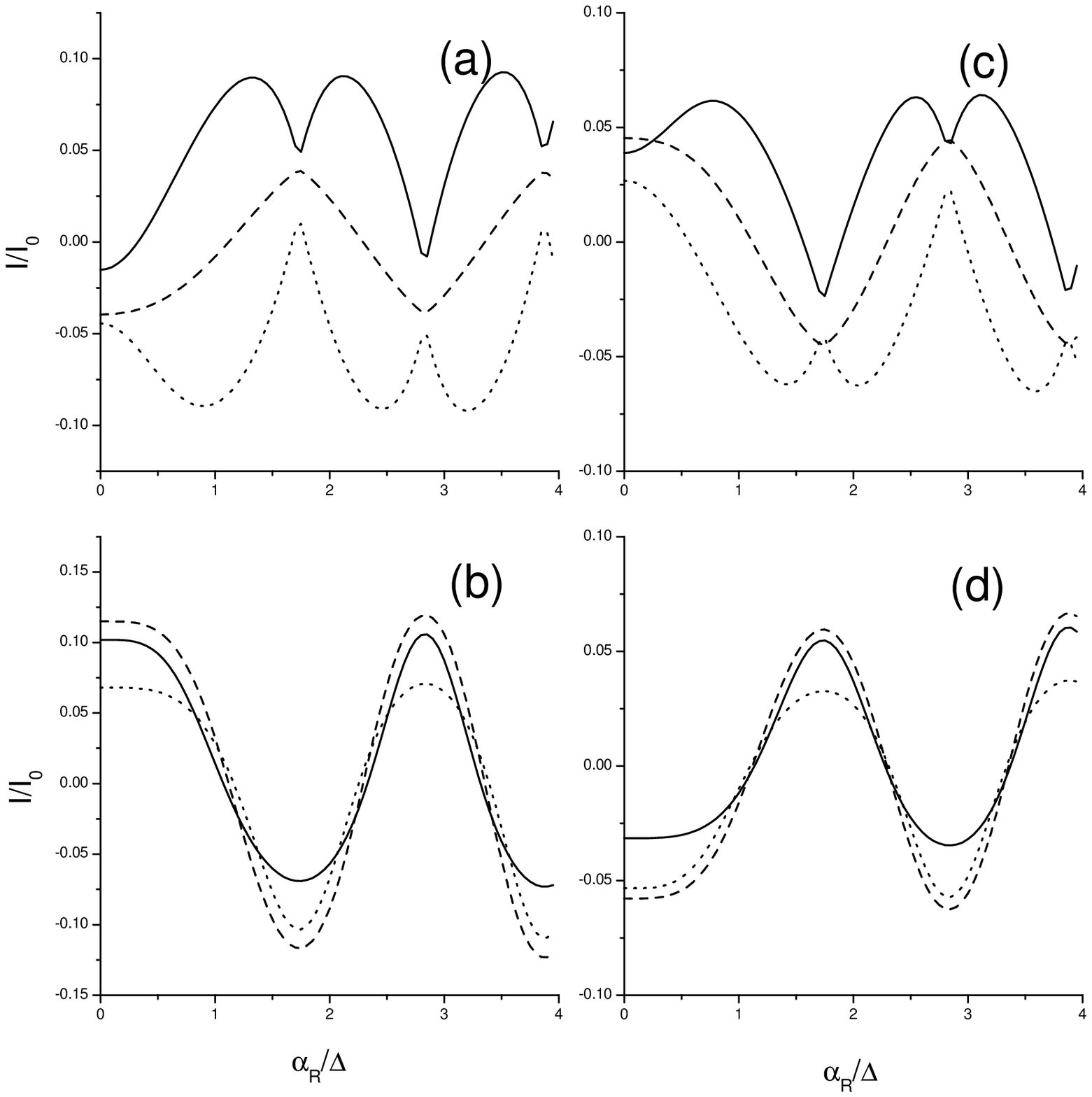}
\caption{ The persistent charge current as a function of the
spin-orbit coupling strength.  The magnetic flux
($\Phi/\Phi_0=0.125$(solid line),$ 0.25$(dashed line),
$0.375$(dotted line)). }
\end{figure}

\begin{figure}[ht]
\includegraphics[width=0.8\columnwidth ]{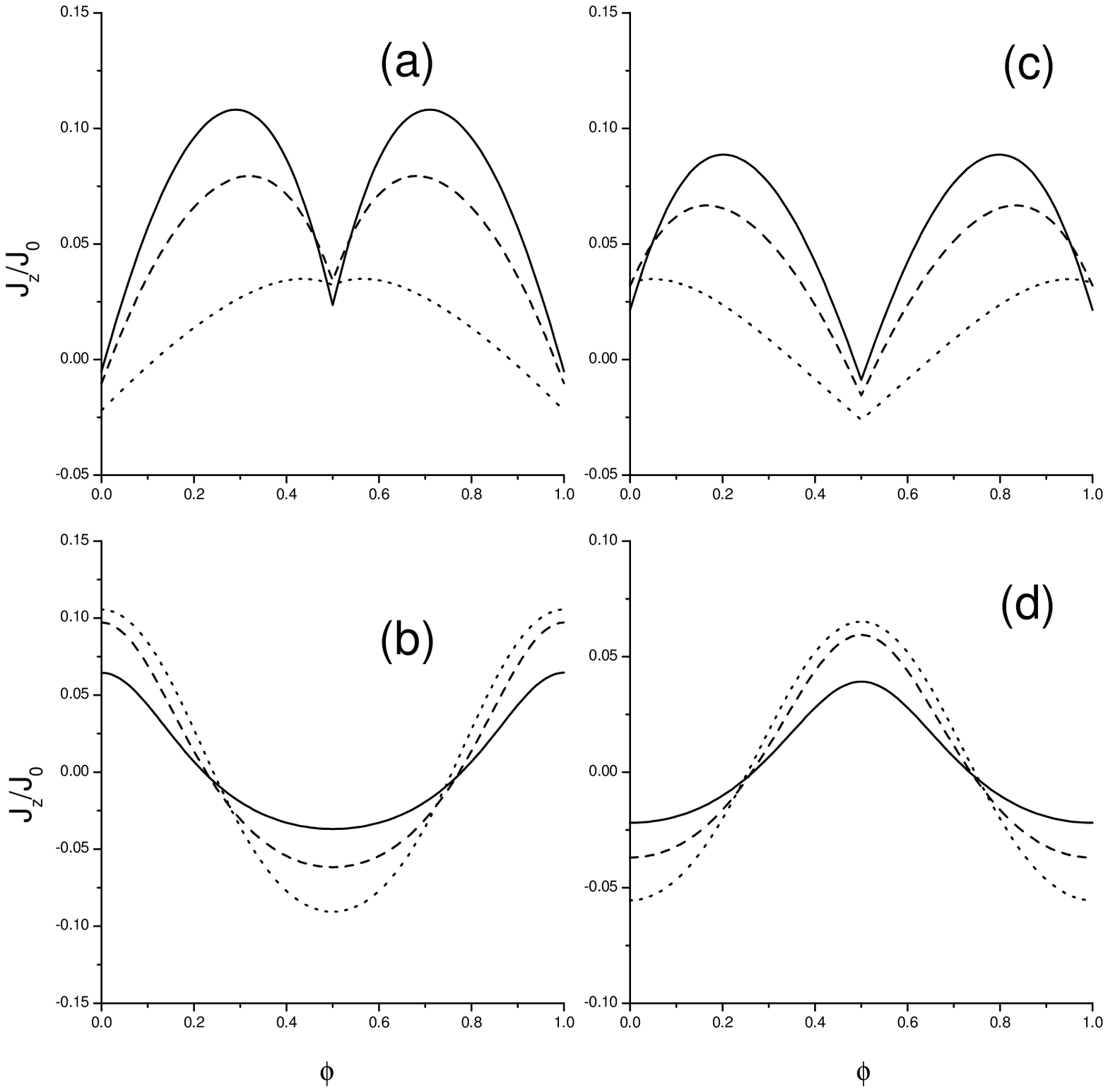}
\caption{ FIG.3: The persistent spin current $J_z$ vs. magnetic
flux for a set of values for the spin-orbit coupling strength(
with $\alpha_R/\Delta=0.5$(solid line),$ 0.7$(dashed line),
$1.0$(dotted line)). The panel (a), (b), (c) and (d) corresponds
the system with total number of electrons $N=99, 100, 101,102$,
respectively. The persistent spin current is measured in units of
$J_0=N \Delta $, and we have taken the other parameter values the
same as that in Fig.1. }
\end{figure}

\begin{figure}[h]
\includegraphics[width=0.8\columnwidth ]{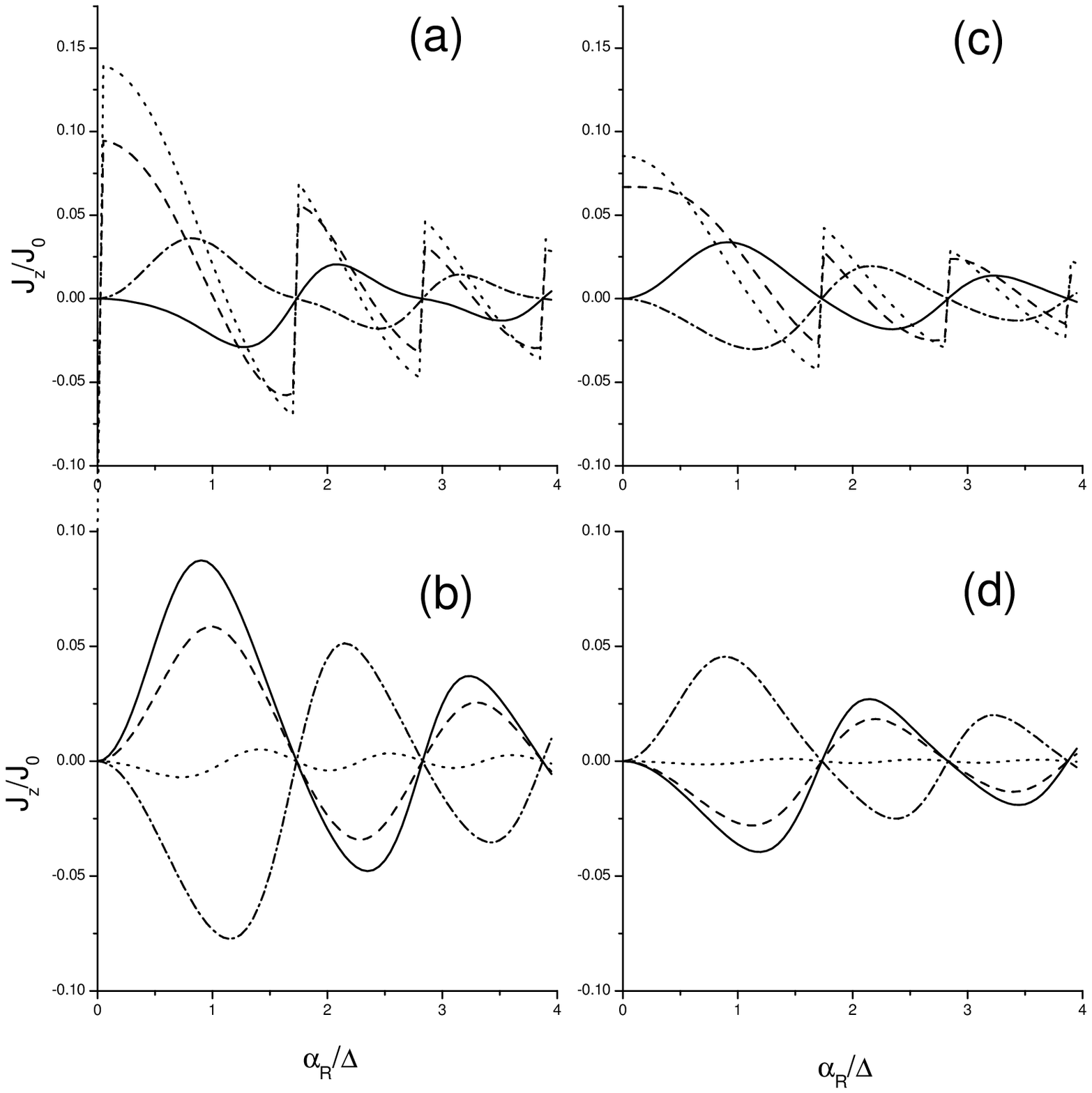}
\caption{ FIG.4: The persistent spin current $J_z$ as a function
of the spin-orbit coupling strength.  The magnetic flux takes the
value ($\Phi/\Phi_0=0.0$(solid line),$ 0.125$(dashed line),
$0.25$(dotted line), $0.5$(dash-dotted line)). }
\end{figure}

\end{document}